\documentclass[12pt,a4paper]{elsart}
\usepackage{latexsym,amssymb,epsfig}
\usepackage{amsmath,eepic}
\newcommand{\di}{\ensuremath{\mathrm{d}}}
\newcommand{\0}{$\phantom{0}$}

\def\0{\phantom{0}}
\renewcommand{\thefootnote}{\fnsymbol{footnote}}
\begin{document}
\begin{center}
{\bf \large A molecular simulation study of shear and bulk }

{\bf \large viscosity and thermal conductivity of simple real fluids}
\bigskip

G.A. Fern\'{a}ndez, J. Vrabec\footnote{To whom correspondence should be addressed, tel.:
+49-711/685-6107, fax: +49-711/685-7657, email: vrabec@itt.uni-stuttgart.de}, and H.
Hasse

Institute of Thermodynamics and Thermal Process Engineering,

University of Stuttgart, D-70550 Stuttgart, Germany

\end{center}

\renewcommand{\thefootnote}{\alph{footnote}}
\baselineskip25pt

\vskip3cm Total number of pages: 31

Number of tables: 2

Number of figures: 7

\bigskip

\clearpage
\textbf{ABSTRACT}

Shear and bulk viscosity and thermal conductivity for argon, krypton, xenon, and methane
and the binary mixtures argon+krypton and argon+methane were determined by equilibrium
molecular dynamics with the Green-Kubo method. The fluids were modeled by spherical
Lennard-Jones pair-potentials with parameters adjusted to experimental vapor
liquid-equilibria data alone. Good agreement between the predictions from simulation and
experimental data is found for shear viscosity and thermal conductivity of the pure
fluids and binary mixtures. The simulation results for the bulk viscosity show only poor
agreement with experimental data for most fluids, despite good agreement with other
simulations data from the literature. This indicates that presently available
experimental data for the bulk viscosity, a property which is difficult to measure, are
inaccurate.


\clearpage

\textbf{1. Introduction}

Transport properties play an important role in many technical and natural processes.
With the rapid increase of available computing power, molecular simulation in
combination with molecular modeling is becoming an interesting option for describing
transport properties in regions where experimental data are not available or difficult
to obtain. The calculation of transport coefficients by molecular simulation can be
achieved by non-equilibrium molecular dynamics (NEMD) or equilibrium molecular dynamics
(EMD). In NEMD, transport coefficients are calculated as the ratio of a flux to an
appropriate driving force, extrapolating to the limit of zero driving force
\cite{evansmorriss}. In EMD transport coefficients are often calculated by the
Green-Kubo formalism \cite{green,kubo}. The choice between EMD and NEMD is largely a
matter of taste and inclination, see e.g. \cite{holian,erpenbeck,cui}. There are
numerous contributions in the literature in which both methods were applied for the
calculation of shear viscosity \cite{holian,pas,rowley,heyes1,heyes2,shoen_sv}, bulk
viscosity \cite{borgelt,heyes3,hoover}, and  thermal conductivity
\cite{pas,heyes1,heyes2,vogel1,vogel2} with comparable performance. To our knowledge the
most comprehensive study on transport coefficients of the spherical Lennard-Jones fluid
is reported in \cite{meier1,meier2}. Despite of the large number of publications on
simulation of transport properties with the spherical Lennard-Jones potential, not much
effort seems to have been spent on a comparison of simulation results with experimental
data of real fluids. Exceptions are the works of Michels et al. \cite{michels} and Heyes
et al. \cite{heyes1,heyes2}. Michels et al. \cite{michels} compared the self-diffusion
coefficient to the Chapman-Enskog theory and experimental data for krypton and methane
at high densities, but thermodynamic properties were not considered. Heyes et al.
\cite{heyes1,heyes2} simulated both transport and thermodynamic properties of
argon+krypton, argon+methane, and methane+nitrogen, but properties of pure fluids were
not considered. In other cases \cite{coelho,stoker} more complex molecules like
ethylene, carbon dioxide, phenol, alkanes, or carbon tetrachloride were modeled by the
spherical Lennard-Jones potential, which surely is an oversimplification. Simulation
data on transport properties were compared there to experimental data, but not the
thermodynamic properties.

The interest here lies in the quantitative evaluation of the performance of
Lennard-Jones models \cite{vrabec1}, which have been optimized for the accurate
prediction of thermodynamic properties, in the description of transport properties. In
this work EMD is used to carry out a comprehensive comparison, of shear and bulk
viscosity and thermal conductivity of pure fluids and binary mixtures of noble gases and
methane. The molecular models are taken from \cite{vrabec1}. These models were adjusted
only to vapor-liquid equilibria and yield accurate descriptions of static thermodynamic
properties over a wide range of temperatures and densities. Furthermore, they were
recently applied for the prediction of diffusion coefficients of pure and binary
mixtures of simple fluids over a wide range of temperatures and densities with good
results \cite{fernandez}.

\clearpage

\textbf{2. Theoretical background}

Transport coefficients are associated to irreversible processes, however, it is possible
to describe irreversible processes in terms of reversible microscopic fluctuations,
through the fluctuation dissipation theory \cite{kubo2}. In that theory, it is shown
that transport coefficients can be calculated as integrals of time-correlation functions
of appropriate quantities \cite{green,kubo}. There are different methods to relate
transport coefficients to time-correlation functions; a good review can be found in
\cite{zwanzig}.

\textit{2.1. Shear and bulk viscosity}

The shear viscosity $\eta_s$, as defined in Newton's "law" of viscosity, describes the
resistance of a fluid to shear forces. It refers to the resistance of an infinitesimal
volume element to shear at constant volume \cite{barton}. The shear viscosity can also
be related to momentum transport under the influence of velocity gradients. From a
microscopic point of view, the shear viscosity can be calculated by integration of the
time-autocorrelation function of the off diagonal elements of the stress tensor
$J_p^{xy}$ \cite{gubbins1,steele}
\smallskip
\begin{equation}\label{shear}
\eta_s=\frac{1}{Vk_BT}\int_{0}^{\infty} \di t~\big\langle J^{xy}_p(t)\cdot
J^{xy}_p(0)\big\rangle,
\end{equation}
where $V$ is the volume, $k_B$ is the Boltzmann constant, $T$ the temperature, and
$<...>$ denotes the ensemble average. The statistics of the ensemble average in Eq.
(\ref{shear}) can be improved using all three independent off diagonal elements of the
stress tensor $J_p^{xy}$, $J_p^{xz}$ and $J_p^{yz}$. For a pure fluid, the component
$J_p^{xy}$ of the microscopic stress tensor $\textbf{J}_p$ is given by
\bigskip
\begin{equation}\label{shear_t}
J_p^{xy}=\sum_{i=1}^N m_i v_{i}^x v_{i}^y -\sum_{i=1}^N \sum_{j>i}^N r_{i j}^x
\frac{\partial u(r_{i j})}{\partial r_{i j}^y}.
\end{equation}
Here $i$ and $j$ are the indices of the particles and the upper indices $x$ and $y$
denote the vector components of the particle velocities $v_i$. Eqs. (\ref{shear}) and
(\ref{shear_t}) can be applied directly to mixtures.

On the other hand, the bulk viscosity $\eta_b$ refers to the resistance to dilatation of
an infinitesimal volume element at constant shape \cite{barton}. The bulk viscosity in
polyatomic molecules, is related to a characteristic time required for the transfer of
energy from the translational to the internal degrees of freedom \cite{HCB}. Moreover,
the bulk viscosity plays an important role to describe ultrasonic wave absorption and
dispersion \cite{karim}. From a microscopic point of view, the bulk viscosity can be
calculated by integration of the time-autocorrelation function of the diagonal elements
of the stress tensor and an additional term that involves the product of pressure $p$
and volume $V$ that does not occur in the shear viscosity, cf. Eq.(\ref{shear}). In the
$NVE$ ensemble the bulk viscosity is given by \cite{gubbins1,steele}
\bigskip
\begin{equation}\label{bulk}
\eta_b=\frac{1}{Vk_BT}\int_{0}^{\infty} \di t~\big\langle (J^{xx}_p(t)-pV(t))\cdot
(J^{xx}_p(0)-pV(0))\big\rangle.
\end{equation}

The component $J_p^{xx}$ of the microscopic stress tensor $\textbf{J}_p$ is given by
\bigskip
\begin{equation}\label{bulk_t}
J_p^{xx}=\sum_{i=1}^N m_i v_{i}^x v_{i}^x -\sum_{i=1}^N \sum_{j>i}^N r_{i j}^x
\frac{\partial u(r_{i j})}{\partial r_{i j}^x}.
\end{equation}

The statistics of the ensemble average in Eq. (\ref{bulk_t}) can be improved using all
three independent diagonal elements of the stress tensor $J_p^{xx}$, $J_p^{yy}$,
$J_p^{zz}$, and their permutations. In the case of mixtures Eqs. (\ref{bulk}) and
(\ref{bulk_t}) can be directly applied. This equation is used in many simulation studies
on the bulk viscosity e.g. \cite{heyes1,heyes2,borgelt,hoover}.

\textit{2.3. Thermal conductivity}

The thermal conductivity $\lambda$, as defined in Fourier's "law" of heat conduction,
characterizes the capability of a substance for molecular transport of energy driven by
temperature gradients. It can be calculated by integration of the time-autocorrelation
function of the elements of the microscopic heat flow $J^x_q$, and is given by
\cite{gubbins1,steele}
\bigskip
\begin{equation}\label{termalc}
\lambda=\frac{1}{Vk_BT^2} \int_{0}^{\infty} \di t~\big\langle J^x_q(t)\cdot J^x_q(0)
\big\rangle.
\end{equation}

Here, the heat flow $\textbf{J}_q$ for a pure fluid is given by
\bigskip
\begin{equation}\label{qflux}
\textbf{J}_q=\frac{1}{2} \sum_{i=1}^N m_iv_i^2 \textbf{v}_i-\sum_{i=1}^N \sum_{j>i}^N
\big[\textbf{r}_{ij} : \frac{\partial u(r_{ij})}{\partial
\textbf{r}_{ij}}-\textbf{I}\cdot u(r_{ij})\big]\cdot \textbf{v}_i,
\end{equation}

where $\textbf{v}_i$ is the velocity vector of particle $i$ and $\textbf{r}_{ij}$ is the
distance vector between particles $i$ and $j$. The term in squared parenthesis denotes
the difference between a dyadic product and the unitary tensor $\textbf{I}$ multiplied
by the intramolecular potential energy $u(r_{ij})$. This description of the heat flow is
not sufficient for binary and multi-component mixtures. In mixtures both diffusion and
energy transport occur coupled \cite{groot}, so that energy can be transported on a
molecular level by diffusion or by heat transport. In a binary mixture with the
components $\alpha$ and $\beta$ the heat flow is given by \cite{vogel1,gubbins1}
\bigskip
\begin{multline}\label{qfluxb}
\textbf{J}_q=\frac{1}{2}\sum_{k=\alpha}^{\beta} \sum_{i=1}^{N_k}m_i^k (v^k_i)^2
\textbf{v}_i^k-\sum_{k=\alpha}^{\beta}\sum_{l=\alpha}^{\beta} \sum_{i=1}^{N_k}
\sum_{j>i}^{N_l}\big[\textbf{r}_{i j}^{kl} : \frac{\partial u(r_{i j}^{kl})}
{\partial{\textbf{r}_{i j}^{kl}}}-\textbf{I}\cdot u(r_{i j}^{kl})\big]\cdot\textbf{v}_i^k\\
-\sum_{k=\alpha}^{\beta} h^k \sum_{i=1}^{N_k} \textbf{v}_i^k,
\end{multline}

where $h^k$ denotes the partial molar enthalpy of component $k$. The computation of the
heat flow in a binary mixture can, in principle, be accomplished in one simulation,
however, here two separate simulations were preferred. One $NpT$ simulation was
performed for the computation of the partial molar enthalpies, corresponding to the
enthalpic part of the energy flow, and another $NVE$ simulation for the calculation of
the autocorrelation function of the heat flow.

\textit{2.4. Molecular models}

In this work, noble gases and methane are considered. These molecules exhibit rather
simple intermolecular interaction so that the description of the molecular interactions
by the Lennard-Jones 12-6 (LJ) potential is sufficient and physically meaningful for
many technically relevant applications \cite{mcdonald}. The LJ potential $u$ is defined
by
\bigskip
\begin{equation}\label{lj126}
u(r_{ij})=4 \epsilon \left[\left(\frac{\sigma} {r_{ij}} \right)^{12}-
\left(\frac{\sigma} {r_{ij}} \right)^6\right],
\end{equation}

where $\sigma$ is the LJ size parameter, $\epsilon$ the LJ energy parameter and $r_{ij}$
the intermolecular distance between particles $i$ and $j$. The parameters $\sigma$ and
$\epsilon$ are taken from \cite{vrabec1} and given in Table \ref{T1}. They were adjusted
to experimental pure substance vapor-liquid equilibrium data alone. For modeling
mixtures, parameters for the unlike interactions are needed. Following previous work of
our group \cite{fernandez,vrabec3,stoll1}, they are given by a modified
Lorentz-Berthelot combination rule with
\bigskip
\begin{equation}\label{berte}
\sigma_{12}=\frac{(\sigma_{11}+\sigma_{22})}{2},
\end{equation}
and
\begin{equation}\label{lorenz}
\epsilon_{12}=\xi \cdot \sqrt{\epsilon_{11}\epsilon_{22}},
\end{equation}
where $\xi$ is an adjustable binary interaction parameter. This parameter allows an
accurate description of the binary mixture data and was determined in previous work by
an adjustment to one experimental bubble point \cite{vrabec3,stoll1}. The binary
interaction parameters used in the present work are listed in Table \ref{T2}.

\textit{2.5. Simulation details}

The molecular simulations were performed in a cubic box of volume $V$ containing $N$=500
or $N$=864 particles modeled by the LJ potential. The cut-off radius was set to $r_{\rm
c}=5 \sigma$, however, for very high densities where $V$ is small, $r_{\rm c}$ was set
to half of the box length, standard techniques for periodic boundary conditions and
long-range corrections were used \cite{allen}. The simulations were started with the
particles in a face-centered-cubic lattice with randomly assigned velocities, the total
momentum of the system was set to zero and Newton's equations of motion were solved with
a Gear predictor-corrector of fifth order \cite{haile}. The time step for this algorithm
was set to $\Delta t \cdot \sqrt{\epsilon_1 / m_1}/\sigma_1=0.001$. All transport
coefficients were calculated in the $NVE$ ensemble, using equations (\ref{shear}),
(\ref{bulk}), (\ref{termalc}) and (\ref{qfluxb}). The relative fluctuations in the total
energy in the $NVE$ ensemble were less than $10^{-4}$ for the longest run. The
simulations were initiated in a $NVT$ ensemble until equilibrium at the desired density
and temperature was reached. Between $100~000$ and $200~000$ time steps were used for
the equilibration depending on the state point. Once the equilibrium is reached, the
thermostat was turned off and then the $NVE$ ensemble invoked to calculate the transport
coefficients by averaging the appropriate autocorrelation function. The length of the
production period depended on density and temperature of the state point. At least
$3~000$ independent autocorrelation functions were used in the calculation of each
coefficient of viscosity and $4~000$ in the calculation of each coefficient of thermal
conductivity. In theory as Eqs. (\ref{shear}), (\ref{bulk}) and \ref{termalc} show, the
value of the transport coefficient are determined by an infinite time integral. In fact,
however, the integral is evaluated based on the length of the simulation. Therefore, the
integration must be stopped at some finite time, ensuring that the contribution of the
long-time tail \cite{alder} is small. Figure 1 shows the behavior of the different
autocorrelation functions and their integrals given by Eqs. (\ref{shear}), (\ref{bulk})
and (\ref{termalc}) for the most dense state points of argon for each transport
property. As can be seen, all autocorrelation functions decay after 2 ps to less than 1
$\%$ of their normalized value. Later they oscillate around zero. To consider the effect
of the long time tail, the calculation of the autocorrelations functions was extended to
5.4 ps for thermal conductivity and shear viscosity and to 6.5 ps for bulk viscosity.
This was done because this autocorrelation function exhibits the largest fluctuation
around zero attributable to long time correlation. The statistical uncertainty of the
transport coefficients and thermodynamic properties were estimated using the Fincham's
method \cite{fincham}. For the calculation of the thermal conductivity of mixtures it is
necessary to include the partial molar enthalpies. For that purpose, Widom's test
particle insertion \cite{sindzingre} was taken using $2~000$ test particles after each
time step, $100~000$ time steps for reaching equilibrium and $300~000$ for production.
Our codes to calculate shear and bulk viscosity and thermal conductivity were
successfully tested with the simulation results of Shoen et al. for viscosity
\cite{shoen_sv}, Heyes \cite{heyes3} for bulk viscosity and Vogelsang et al.
\cite{vogel1} for thermal conductivity. \clearpage

\textbf{3. Results and Discussion}

In this section the prediction of shear and bulk viscosity and thermal conductivity are
compared pointwise with experimental data. For the shear viscosity the correlation of
Rowley et al. \cite{rowley,rowley2}, which is based on molecular simulation results, was
also used.

Figure 2 shows the results for the shear viscosity of argon, krypton, xenon and methane
in comparison with experimental data. The data are reported at different temperatures
and were taken from Vargaftik \cite{vargaf} for the noble gases and from Evers et {\it
al}. \cite{evers} for methane. Overall, very good agreement between simulation and
experimental data is found. The lowest relative deviations are found for argon and
krypton with a few percent at lower densities. Also the results of shear viscosity of
xenon and methane show very good agreement at low density, however, as the density
increases, the deviations from the experimental data reach up to about $15\%$ for xenon
and about $20\%$ for methane. It can be observed that the simulations for krypton, xenon
and methane tend to underestimate the experimental viscosities as the density increases.
This underestimation is larger in the results given by Rowley's correlation for the
shear viscosity.

Figure 3 shows the results for the shear viscosity of the binary mixtures argon+krypton
and argon+methane for two temperatures, the experimental data were taken from
Mikhailenko et {\it al}. \cite{mikha,mikha2}. Good agreement between simulation and
experimental data is found. For the mixture argon+krypton the typical deviations are
about $10\%$ and the highest deviations occur for krypton-rich state points. The
predictions for the mixture argon+methane show a better agreement with the experimental
data than those for argon+krypton. Typical deviations of the mixture argon+methane are
about $10\%$ at 100 K and $6\%$ at 120 K, simulations performed at one intermediate
temperature ($T$=110 K) confirm that better agreement between simulation and experiments
is found as the temperature increases. The comparison of the present simulations with
previous simulations of Heyes \cite{heyes1,heyes2}, shows a comparable agreement for the
mixture argon+krypton, however, in the mixture argon+methane the present simulations
show better agreement than those of Heyes, specially in methane-rich state points, c.f.
Fig. 3.

Figure 4 shows the results for the bulk viscosity of argon, krypton, xenon and methane
in comparison with experimental data. The experimental data are reported at different
temperatures and were taken from Cowan et {\it al}., Malbrunotet et {\it al}., Cowan et
{\it al}. and Singer \cite{cowan,malbrunot,cowan2,singer}, respectively. The agreement
is poor. Neither the density dependence nor the absolute value of $\eta_b$ predicted by
molecular simulation agrees with the experimental data. The best results for $\eta_b$
are achieved at the low temperatures and high densities for krypton. In this case the
typical error is about $13 \%$, even here the density dependence in not predicted
correctly. For the other fluids, the predictions are lower than the experimental data by
about $50 \%$. Likewise, the experimental data of bulk viscosity show a stronger
dependence on the density than the simulations. It must be pointed out here, however,
that the method to measure the bulk viscosity by means of acoustic absorption of sound
waves, involves considerable error \cite{graves}. Among the quoted experimental data,
the krypton data are claimed to be the most accurate with an error band of about $25
\%$, for the remainder error bands of up to $40 \%$ can be assumed. In the light of the
fact that these simple molecular models describe both the thermal and caloric properties
accurately \cite{vrabec2} as well as the transport properties self-diffusion
\cite{fernandez}, shear viscosity and thermal conductivity (see below), it can be argued
that the deviations founded for the bulk viscosity are due to experimental error.

Figure 5 shows the results for bulk viscosity of the binary mixtures argon+ krypton and
argon+methane. The agreement is again poor. Neither the composition dependence nor the
absolute value of $\eta_b$ predicted by molecular simulation agrees with the
experimental data. The best results for $\eta_b$ are achieved at the lowest temperature
for the mixture argon+krypton. In this case, a discrepancy of about $50 \%$ is found. In
agreement with the results for pure fluids, it is found that the predictions are much
too low in comparison to experimental data. Previous work on these mixtures by Heynes et
{\it al}. \cite{heyes1,heyes2} confirms lower values from simulation, c.f. Fig. 5.

Figure 6 shows the results for the thermal conductivity of argon, krypton, xenon and
methane in comparison with experimental data. The data are reported at different
temperatures along the bubble line and were taken from Vargaftik et {\it al}.
\cite{vargaf}. Overall, very good agreement between simulation and experimental data is
present. The lowest relative deviations are found for argon and xenon, typical values
are $4\%$ for argon and $7\%$ for xenon. The deviations for krypton and methane reach up
to about $20\%$, no tendency is observed in these deviations. The good agreement for
methane is especially remarkable, considering that the molecular model is very
simplified and does not consider the contribution of rotation or internal degrees of
freedom \cite{evans,wang,tokumasu}.

Figure 7 shows the results for the thermal conductivity of the binary mixtures
argon+krypton and argon+methane for two temperatures, the experimental data were taken
from Mikhailenko et {\it al}. \cite{mikha,mikha2}. In general good agreement between
simulation and experimental data is found. Due to the error introduced by the partial
molar enthalpy, the statistical uncertainty of the thermal conductivity was estimated as
$5 \%$. For the mixture argon+krypton the typical deviations are about $5\%$ at 120 K,
and about $7\%$ at 140 K. For most of the simulated state points, these deviations lie
within the uncertainty bars. For the mixture argon+methane a better agreement is found
than for the mixture argon+krypton. Over the whole composition range, simulation and
experiment for argon+methane agree within the statistical uncertainties. The comparison
of the present simulations with previous simulations of Heynes \cite{heyes1,heyes2}
shows good agreement, c.f. Fig. 7. \clearpage

\textbf{4. Conclusion}

In the present work, the Green-Kubo formalism was used  to calculate transport
properties for pure and binary mixtures of four noble gases and methane. The molecular
interactions of the fluids were modeled by the spherical Lennard-Jones pair potential
with parameters adjusted to vapor-liquid equilibrium only. A comprehensive comparison
with available experimental data shows good agreement for pure fluids and binary
mixtures for shear viscosity and thermal conductivity. On the other hand, for the bulk
viscosity, with the exception of pure krypton, considerable systematic deviations
between simulations and experiment occur. This disagreement hints towards highly
inaccurate measurements. The present results support the finding that the spherical LJ
12-6 potential is an adequate description for the regarded noble gases and also methane,
in spite of the simplicity of the used model. Likewise the modified Lorentz-Berthelot
combination rules with one binary interaction parameter are an adequate description of
the molecular binary unlike interaction. It is worthwhile to extend the study to more
complex fluids.\clearpage

\textit{List of Symbols}

\begin{tabular}{ll}
$E$ & Energy \\
$h^k$ & partial molar enthalpy of component $k$ \\
$i$ & particle counting index \\
$j$ & particle counting index \\
$J_p^{xy}$ & $xy$ stress tensor element\\
$J_q^x$ & $x$ heat flow element\\
$k_B$ & Boltzmann constant \\
$k$ & species counting index \\
$M$ & molar mass \\
$m$ & molecular mass \\
$N$ & number of particles \\
$r$ & intermolecular distance \\
$r_{\rm c}$ & cut-off radius \\
$t$ & time \\
$T$ & temperature \\
$u$ & pair potential energy \\
$v$ & velocity \\
$V$ & volume \\
$x$ & cartesian coordinate  \\
$y$ & cartesian coordinate  \\
$z$ & cartesian coordinate  \\
\end{tabular}
\clearpage
\textit{Greek Symbols}

\begin{tabular}{ll}
$\alpha$ & component\\
$\beta$ & component\\
$\Delta t$ & integration time step\\
$\epsilon$ & Lennard-Jones energy parameter\\
$\eta_s$ & shear viscosity\\
$\eta_v$ & bulk viscosity\\
$\lambda$ & thermal conductivity\\
$\xi$ & adjustable binary interaction parameter\\
$\sigma$ & Lennard-Jones size parameter \\

\end{tabular}

\textit{Vectorial and tensorial quantities}

\begin{tabular}{ll}
$\textbf{I}$ & unitary matrix\\
$\textbf{J}_p$ & stress tensor\\
$\textbf{J}_q$ & heat flow vector\\
$\textbf{r}_{ij}$ & distance vector \\
$\textbf{v}_i$ & velocity of particle $i$\\

\end{tabular}

\clearpage


\begin{table}[t]
\noindent \caption{Potential model parameters for the pure fluids used in this work
\cite{vrabec1} and molar mass \cite{poling}.} \label{T1}
\bigskip
\begin{center}
\begin{tabular}{|l||c|c|c|c|c|} \hline
Fluid    &  \0\0$\sigma$ / \r{A}\0\0   &  $( \epsilon /k_{\rm B}$) / K  & $M$ / g/mol \\\hline\hline
 $\rm  neon $    & 2.8010         & \033.921       &  \020.180                                   \\ \hline
 $\rm  argon $   & 3.3952         &  116.79\0      &  \039.948                                   \\ \hline
 $\rm  krypton $ & 3.6274         &  162.58\0      &  \083.8\0\0                               \\ \hline
 $\rm  xenon   $ & 3.9011         &  227.55\0      &   131.29\0                                   \\ \hline
 $\rm  methane $ & 3.7281         &  148.55\0      &  \016.043                                    \\ \hline
\end{tabular}

\end{center}
\end{table}
\clearpage

\begin{table}[t]
\noindent \caption{Binary interaction parameters taken from \cite{vrabec3}.} \label{T2}
\bigskip
\begin{center}
\begin{tabular}{|l|c|} \hline
\0 Mixture \0     & \0\0 $\xi$ \0\0  \\ \hline\hline
 $\rm argon+ \rm krypton$     & 0.988    \\ \hline
 $\rm argon+ \rm methane$     & 0.964    \\ \hline
\end{tabular}
\end{center}
\end{table}
\clearpage \clearpage

\clearpage
\listoffigures
\clearpage
\begin{figure}[ht]
\caption[Large plots: Autocorrelation functions. Small plots: Integrals following Eqs.
(\ref{shear}), (\ref{bulk}) and (\ref{termalc}). All plots are shown for the most dense
state points of argon for each transport property; top: thermal conductivity $T$=90 K
and $\rho$=34433 mol $\cdot$ m$^{-3}$, middle: shear viscosity $T$=150.7 K and
$\rho$=35046 mol $\cdot$ m$^{-3}$, bottom: bulk viscosity $T$=100 K and $\rho$=32843 mol
$\cdot$ m$^{-3}$. ]{}\label{fig1a}
\begin{center}
\includegraphics[width=150mm,height=200mm]{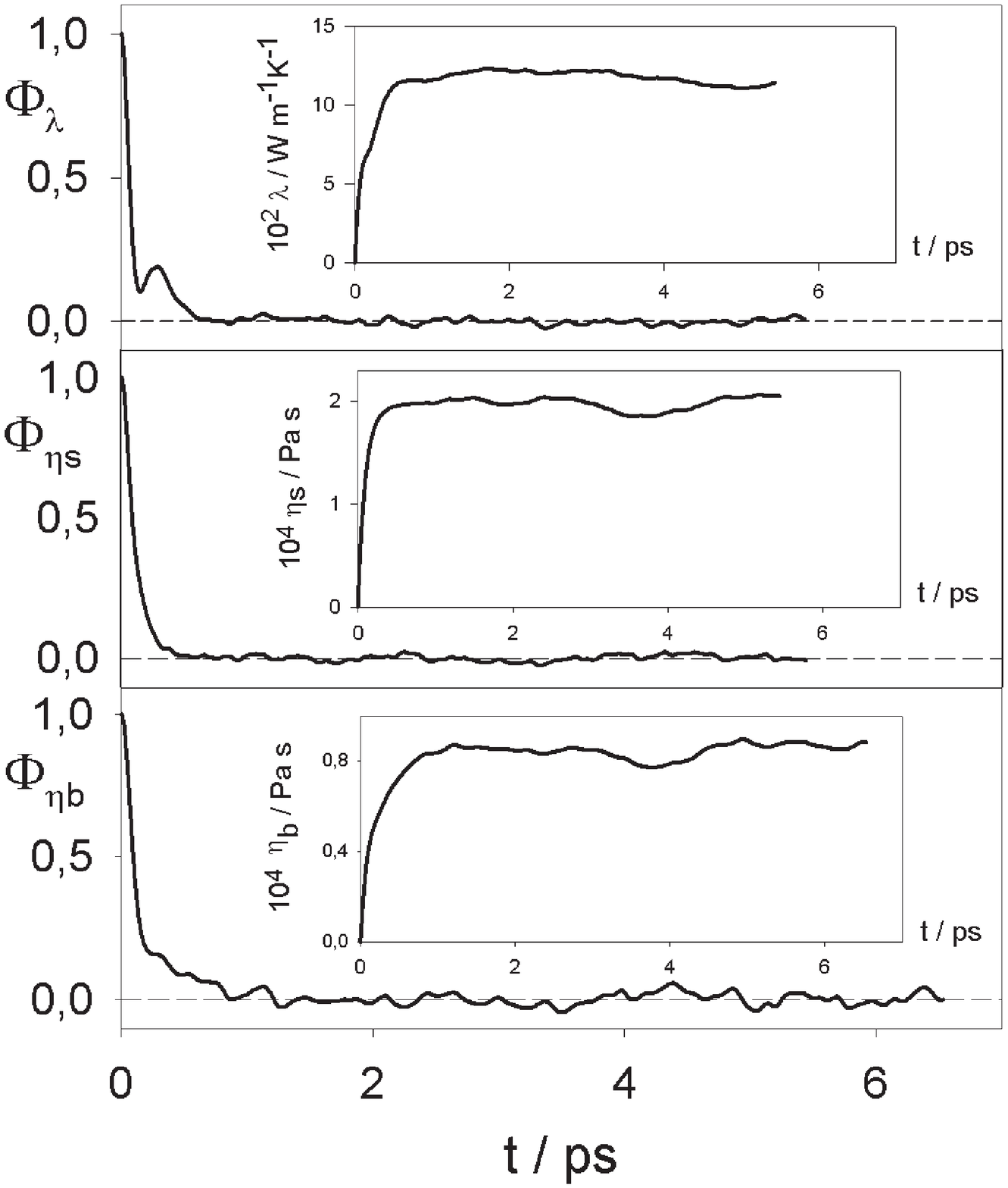}
\end{center}
\end{figure}

\begin{figure}[ht]
\caption[Shear viscosity of argon, krypton, xenon and methane predicted by molecular
simulation (full symbols) compared to experimental data (empty symbols)
\cite{vargaf,evers}. argon $T$=300~K {$\blacktriangle$}; krypton $T$=230~K
{$\blacksquare$}; xenon $T$=270~K {$\blacktriangledown$}; methane $T$=100~K~-~293.15~K
{$\blacklozenge$}; correlation of Rowley et al. \cite{rowley} {\large $-$}
.]{}\label{fig1}
\begin{center}
\includegraphics[width=150mm,height=200mm]{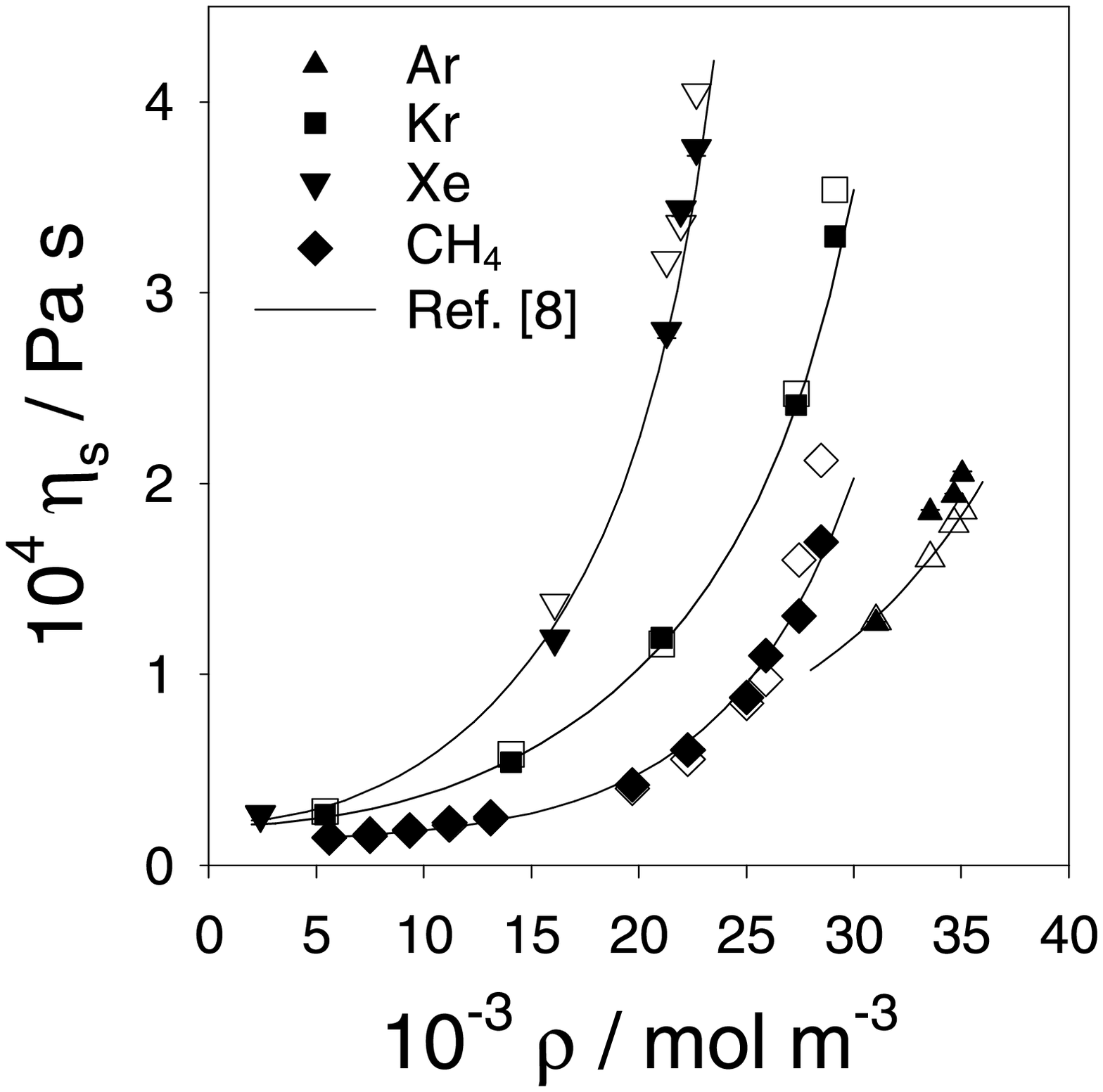}
\end{center}
\end{figure}

\begin{figure}[ht]
\caption[Shear viscosity of the mixtures argon+krypton (top) and argon+metane (bottom)
predicted by molecular simulation (full symbols) compared to experimental data (empty
symbols) \cite{mikha,mikha2}. The lines are a guide for the eye.]{}\label{fig2}
\begin{center}
\includegraphics[width=150mm,height=200mm]{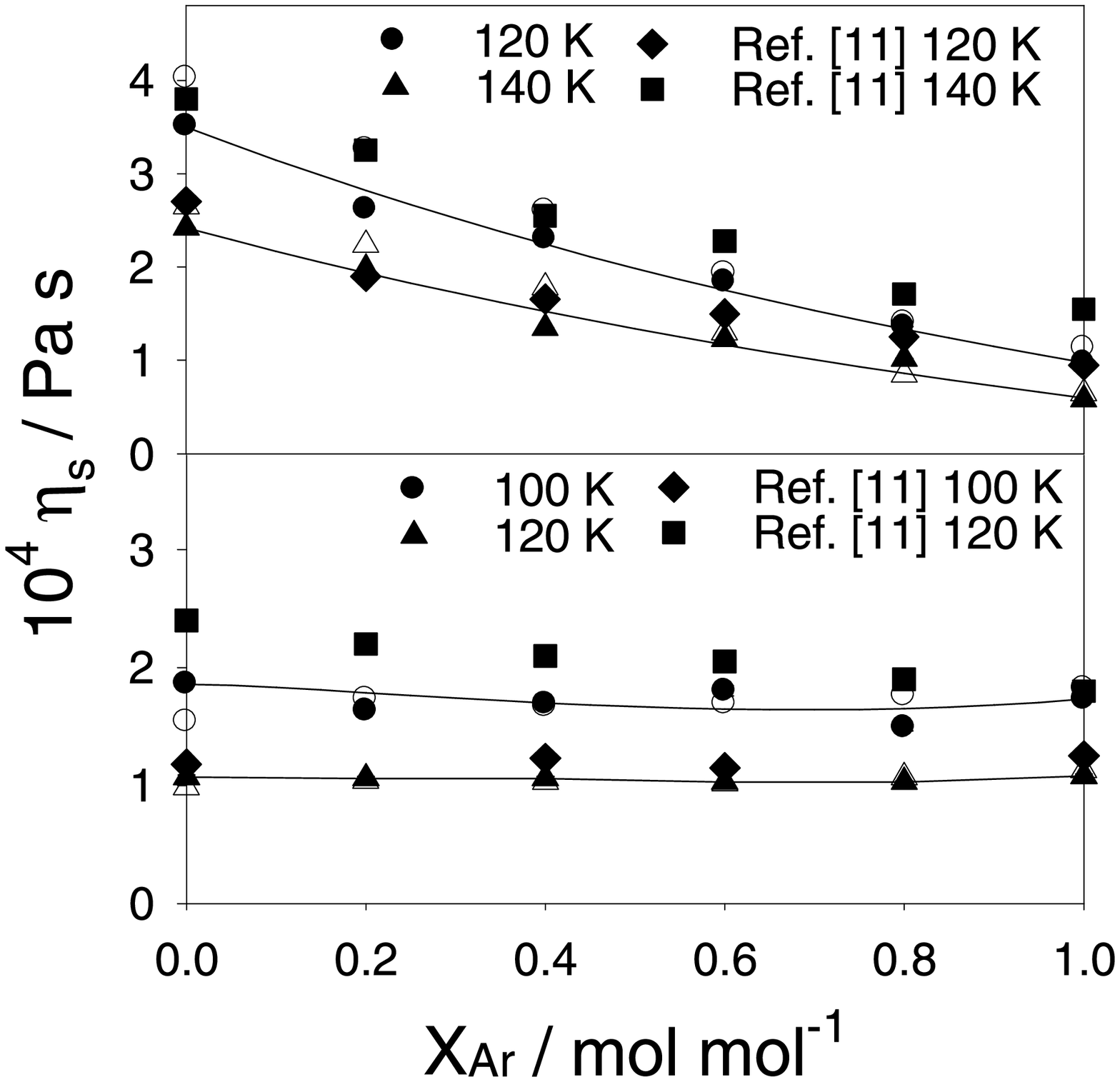}
\end{center}
\end{figure}

\begin{figure}[ht]
\caption[Bulk viscosity of argon, krypton, xenon and methane predicted by molecular
simulation (full symbols) compared to experimental data (empty symbols)
\cite{cowan,malbrunot,cowan2,singer}. The lines are a guide for the eye. argon
$T$=100~K~-~145~K {$\blacktriangle$}; krypton $T$=116~K~-~130~K {$\blacksquare$}; xenon
$T$=165~K~-~265~K {$\blacktriangledown$}; methane $T$=100~K~-~293.15~K
{$\blacklozenge$}.]{}\label{fig3}
\begin{center}
\includegraphics[width=150mm,height=200mm]{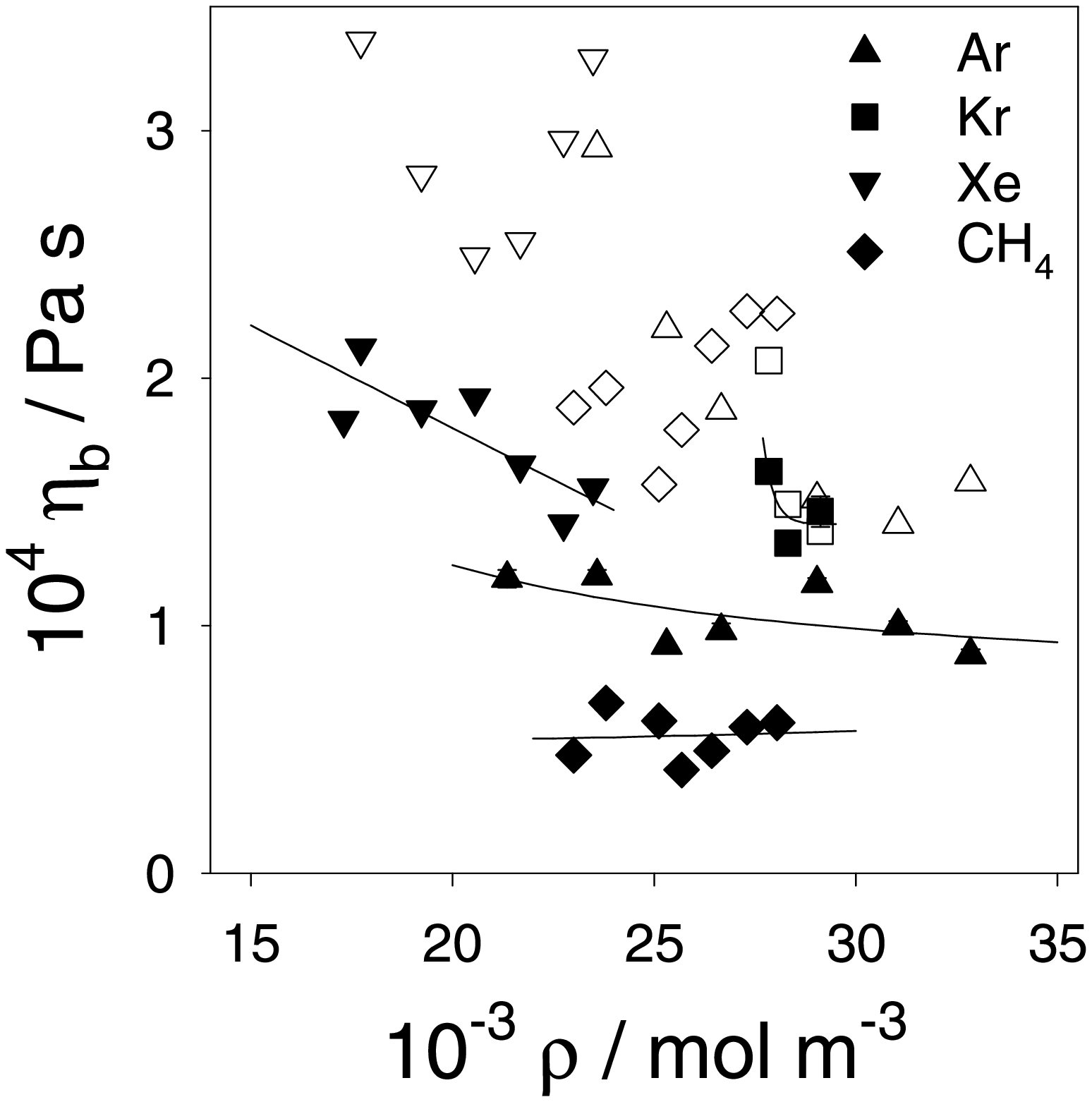}
\end{center}
\end{figure}

\begin{figure}[ht]
\caption[Bulk viscosity of the mixtures argon+krypton (top) and argon+metane (botton)
predicted by molecular simulation (full symbols) compared to experimental data (empty
symbols) \cite{mikha,mikha2}. The lines are a guide for the eye.]{}\label{fig4}
\begin{center}
\includegraphics[width=150mm,height=200mm]{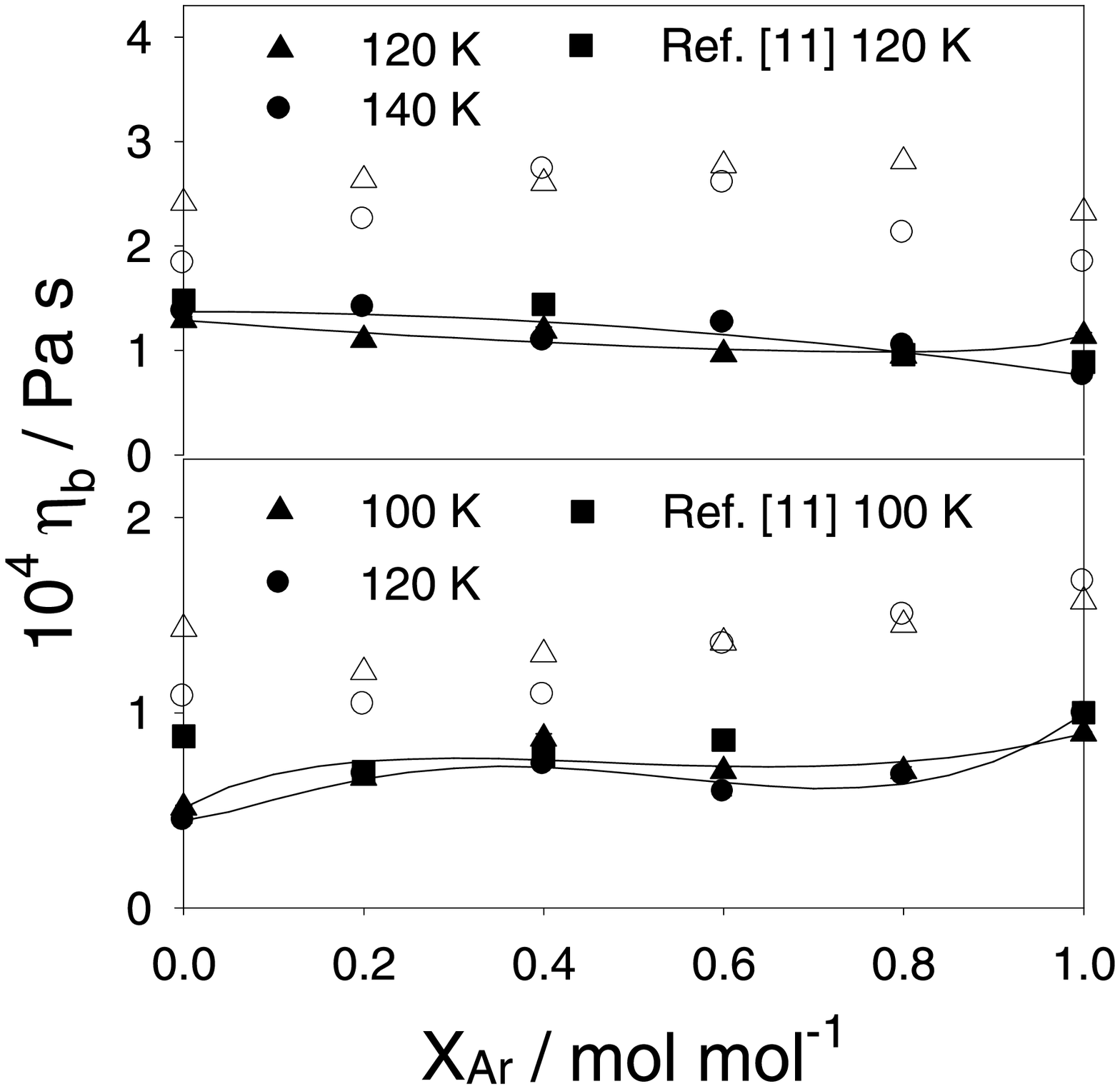}
\end{center}
\end{figure}

\begin{figure}[ht]
\caption[Thermal conductivity of argon, krypton, xenon and methane predicted by molecular
simulation (full symbols) compared to experimental data (empty symbols) \cite{vargaf}.
The data correspond to bubble points reported at different temperatures. argon
$T$=90~K~-~140~K {$\blacktriangle$}; krypton $T$=140~K~-~184~K {$\blacksquare$}; xenon
$T$=170~K~-~270~K {$\blacktriangledown$}; methane $T$=100~K~-~180~K
{$\blacklozenge$}.]{}\label{fig5}
\begin{center}
\includegraphics[width=150mm,height=200mm]{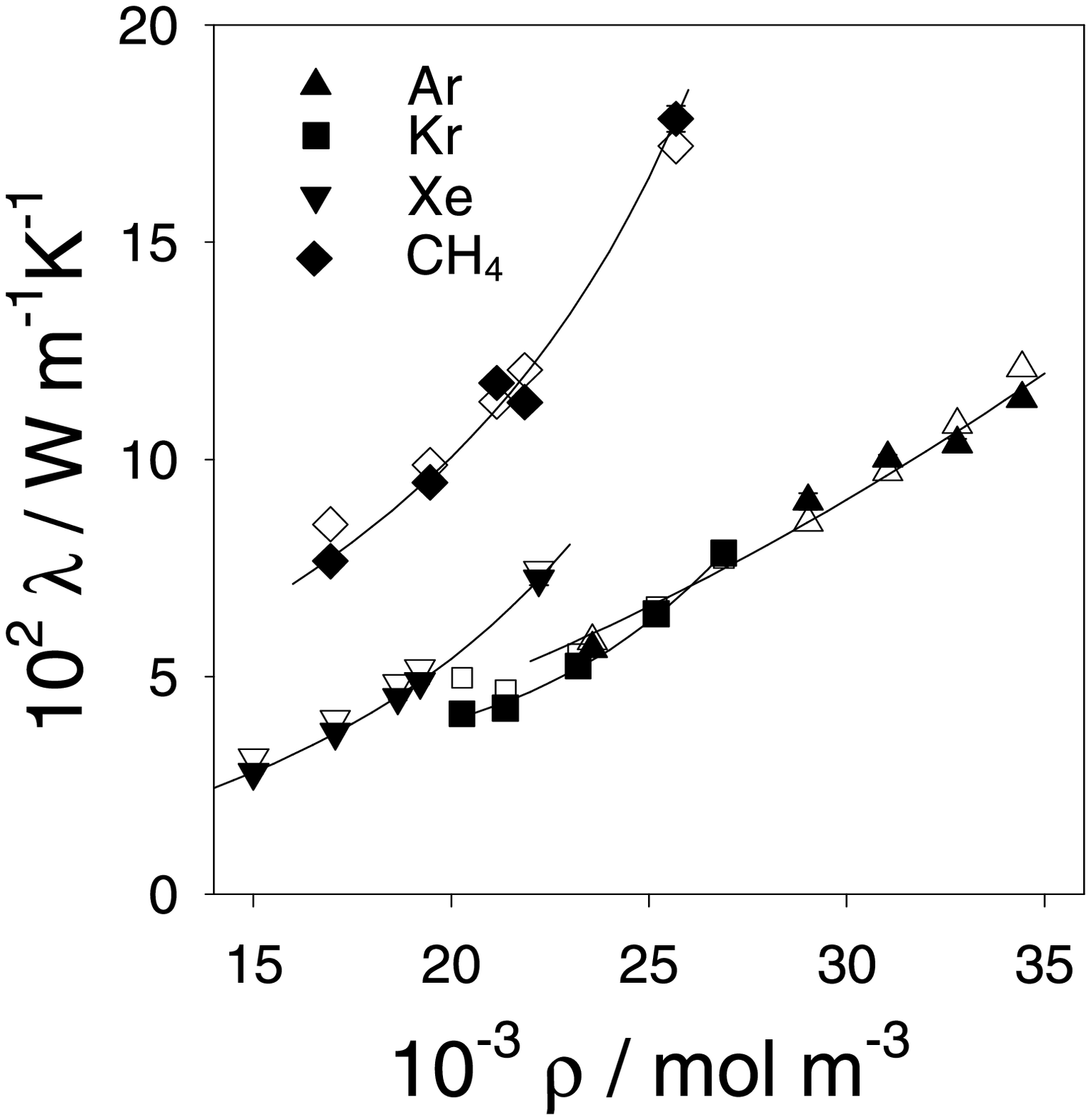}
\end{center}
\end{figure}

\begin{figure}[ht]
\caption[Thermal conductivity of the mixtures argon+krypton (top) and argon+metane
(botton) predicted by molecular simulation (full symbols) compared to experimental data
(empty symbols) \cite{mikha,mikha2}. The lines are a guide for the eye.]{}\label{fig6}
\begin{center}
\includegraphics[width=150mm,height=200mm]{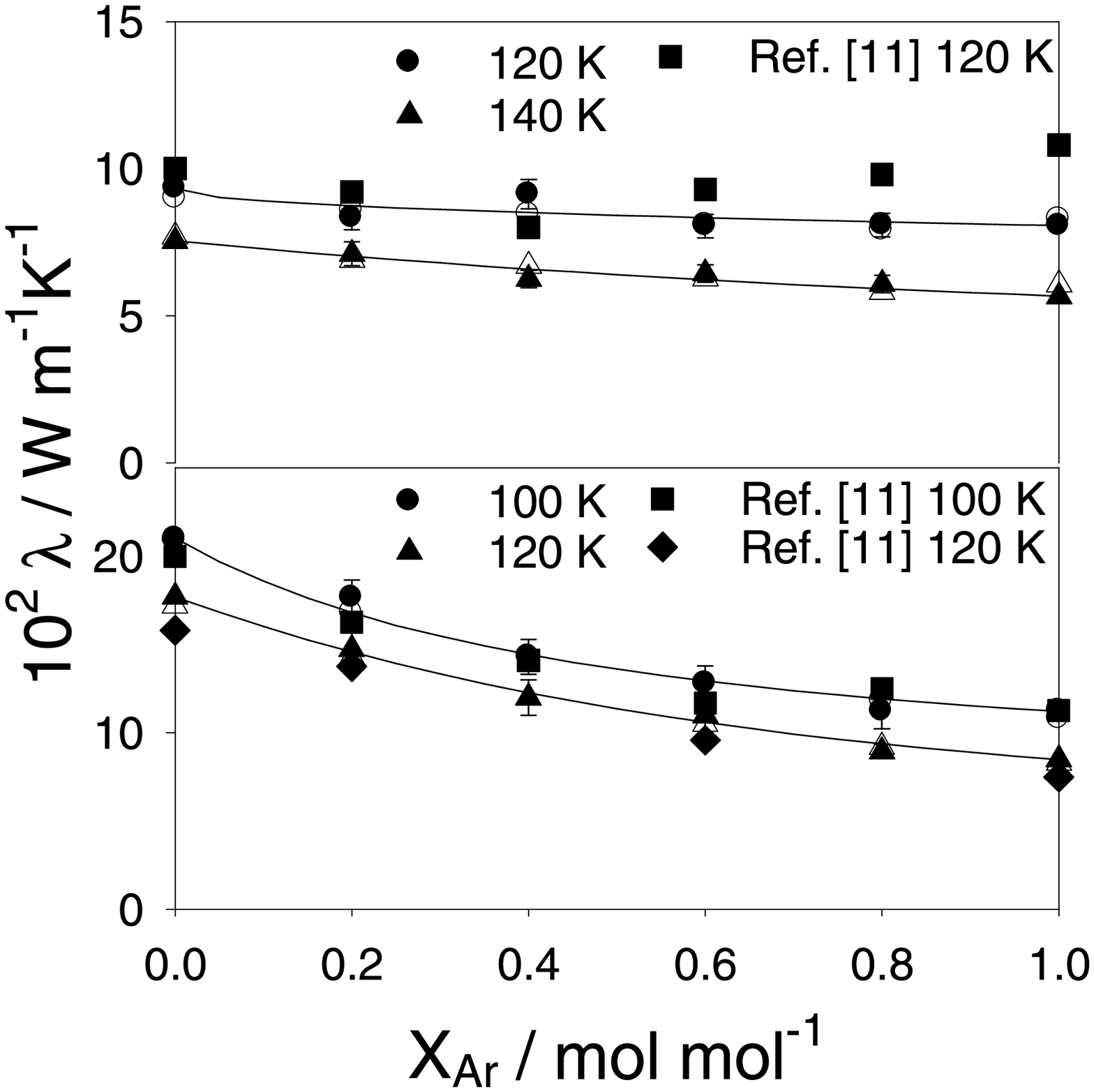}
\end{center}
\end{figure}

\end{document}